\documentclass[journal,12pt,onecolumn,draftclsnofoot]{IEEEtran}
\usepackage[latin1]{inputenc}
\usepackage{times,amsmath}
\usepackage{amsfonts}
\usepackage{pstool}
\usepackage{subfigure}
\usepackage{multirow}
\usepackage{enumerate}
\usepackage{graphicx}
\usepackage{MnSymbol}
\usepackage{stfloats}
\usepackage[table]{xcolor}
\usepackage[square, comma, sort&compress, numbers]{natbib}
\usepackage{nohyperref}
\usepackage{algorithm,algorithmic}
\usepackage{bigints}
\usepackage{amsmath}

\newcounter{tempEquationCounter}
\newcounter{thisEquationNumber}

\makeatletter
\newcommand{\vast}{\bBigg@{4}}
\newcommand{\Vast}{\bBigg@{5}}
\makeatother

\graphicspath{ {Figures/} }

\begin{document}

\title{On the Impact of Mode Selection on Effective Capacity of Device-to-Device Communication}

\author{
\IEEEauthorblockN{S.\ Waqas\ H.\ Shah, M.\ Mahboob\ Ur\ Rahman, Adnan\ N.\ Mian, Ali\ Imran, \\ Shahid\ Mumtaz, and Octavia\ A.\ Dobre}
}

\maketitle

\long\def\symbolfootnote[#1]#2{\begingroup%
\def\thefootnote{\fnsymbol{footnote}}\footnote[#1]{#2}\endgroup}
\symbolfootnote[0]{\hrulefill \\
Syed Waqas Haider Shah, Muhammad Mahboob Ur Rahman, and Adnan Noor Mian are with the Electrical engineering Department, Information Technology University, Lahore 54000, Pakistan (\{waqas.haider, mahboob.rahman, adnan.noor\}@itu.edu.pk). Adnan Noor Mian is also with the Computer Lab, 15 JJ Thomson Avenue, University of Cambridge, UK. \\
Ali Imran is with the Department of Telecommunications Engineering, University of Oklahoma-Tulsa, USA (ali.imran@ou.edu). \\
Shahid Mumtaz is with the Instituto de Telecommuninicoes, DETI, Universidade
de Aveiro, Aveiro 4554, Portugal (smumtaz@av.it.pt). \\
Octavia A. Dobre is with the Department of Electrical and Computer Engineering, Memorial University, St.  John's, NL A1B 3X5, Canada (odobre@mun.ca). \\
This work is partially supported by the National Science Foundation under Grant Numbers 1559483, 1619346, 1718956 and 1730650, as well as the Natural Sciences and Engineering Research Council of Canada through its Discovery program.
}

\maketitle

\begin{abstract}

Consider a device-to-device (D2D) link which utilizes the mode selection to decide between the direct mode and cellular mode. This paper investigates the impact of mode selection on effective capacity (EC)--the maximum sustainable constant arrival rate at a transmitter's queue under statistical quality-of-service constraints--of a D2D link for both overlay and underlay scenarios. Due to lack of channel state information, the transmit device sends data at a fixed rate and fixed power; this fact combined with mode selection makes the D2D channel a Markov service process. Thus, the EC is obtained by calculating the entries of the transition probability matrix corresponding to the Markov D2D channel. Numerical results show that the EC decays exponentially (and the gain of overlay D2D over underlay D2D diminishes) with the increase in estimation error of the pathloss measurements utilized by the mode selection.

\end{abstract}


\section{Introduction}
\label{sec:intro}

Device-to-device (D2D) communication is a regime where a transmit user equipment (UE) talks to the corresponding receive UE (in the close proximity) directly, without routing its data through the base station/eNodeB. D2D communication can alleviate the spectrum scarcity problem by increasing the spectrum reuse, helps realize energy-efficient (due to reduced transmission power levels) and low-latency (because of the direct link) systems, and leads to high data rates. Therefore, D2D is poised to become one of the enabling technologies for upcoming 5G cellular networks \cite{lin2014overview}, \cite{Octavia:WC:2017}.

D2D has attracted significant attention recently. However, due to space constraint, only a few selected and relevant works have been summarized. In \cite{yang2017transmission}, the authors study the impact of user density, transmit power, and link distance on the capacity of cooperative (relay-assisted) D2D links. \cite{AImran:CommL:2015} computes the average coverage probability of a cellular user, in the presence of a number of potential D2D pairs. The authors of \cite{AImran:CST:2016} discuss the separation of the control and data planes to enable D2D communication for 5G networks. In \cite{mahmood2013mode}, Mahmood et. al. study the mode selection problem whereby a D2D link decides between the direct mode and cellular mode.

The effective capacity (EC), on the other hand, is the maximum sustainable {\it constant} arrival rate at a transmitter (queue) in the face of a randomly time-varying (channel) service, under quality-of-service (QoS) constraints \cite{Wu:TWC:2003}. EC has attracted significant attention; to date, EC-based QoS-constrained performance analysis has been carried out for: cognitive radio channels \cite{Gursoy:TWC:2010}, \cite{Anwar:TVT:2016}, systems with various degrees of channel knowledge at the transmitter \cite{gross2012scheduling}, two-hop systems \cite{Gursoy:TIT:2013}, \cite{Lateef:TC:2009}, and correlated fading channels \cite{Soret:TWC:2010}.

Very recently, a couple of works on EC analysis of D2D have emerged. For a given QoS constraint, \cite{Cheng:JSAC:2016} proposes various optimal and sub-optimal power allocation schemes to maximize the EC for both D2D and cellular links operating in underlay mode and overlay mode, respectively. \cite{Mi:ICT:2016} extends the work in \cite{Cheng:JSAC:2016} by performing the EC analysis when the cellular users and D2D users have different QoS requirements. Finally, \cite{Ismaiel:TVT:2018} performs the EC analysis for a D2D link that is used to offload the WiFi traffic. {\it However, to the best of the authors' knowledge, the impact of mode selection on EC of the D2D communication has not been studied in the literature}.

{\bf Outline.} Section II introduces the system model and the EC concept. Section III describes a feature-based mode selection method. The impact of mode selection on EC of overlay and underlay D2D is analyzed in Section IV. Section V provides numerical results. Section VI concludes the paper.

\section{System Model and Background}
\label{sec:sys-model}

\subsection{System Model}
\begin{figure}[ht]
\begin{center}
	\includegraphics{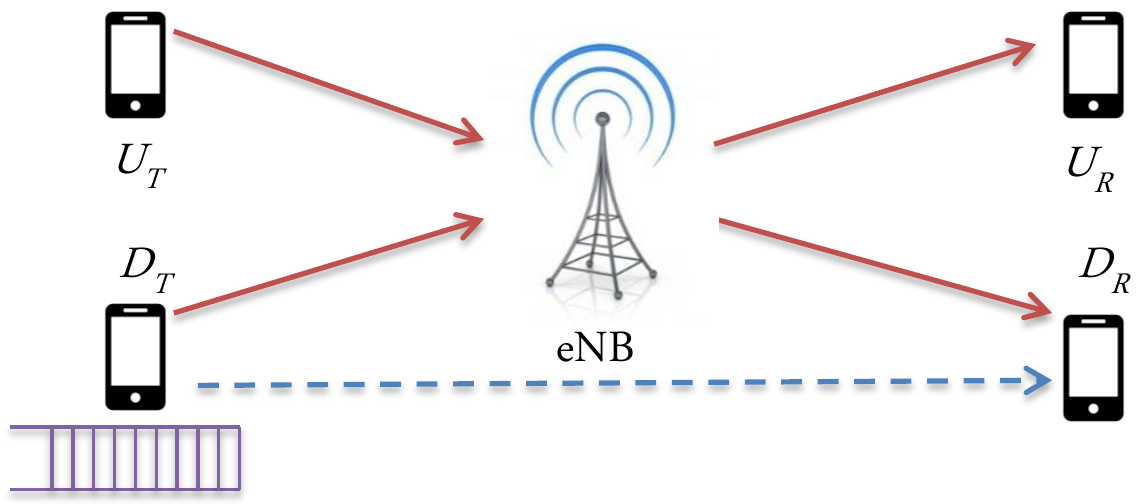}
\caption{The system model: the red arrows represent the cellular links, while the blue arrows represent the potential D2D link. }
\label{fig:sm}
\end{center}
\end{figure}

Consider the coverage region of a single cell consisting of two pairs of UEs forming two communication links (see Fig. \ref{fig:sm}). The first pair of UEs operates in cellular mode, i.e., the transmit UE ($U_T$) communicates with the receive UE ($U_R$) via the eNodeB (eNB). We term the second pair of UEs as D2D candidate pair, whereby the transmit UE ($D_T$) could communicate with the receive UE ($D_R$) either directly (direct mode) or via the eNB (cellular mode). This problem of selecting between the direct mode and cellular mode (at $D_T$) during every time-slot is known as mode selection \cite{Octavia:WC:2017}.

The D2D communication consists of two distinct scenarios: overlay and underlay. In the former, the eNB allocates orthogonal resource blocks (i.e., sub-carriers, time-slots) to the D2D pair and the cellular pair, while in the latter, the D2D pair reuses the resource blocks of the cellular pair.

We further assume the following: i) all channels are block-fading channels with Rayleigh distribution; ii) the eNB employs decode-and-forward (DF) strategy to relay the transmission of an UE to the corresponding receive UE when the pair of UEs operates in cellular mode; iii) the system is time-slotted with $\tau$ second long time-slots.

\subsection{Background: Effective Capacity (EC)}
The EC is defined as the log moment generating function (MGF) of the cumulative service process $S(t)$ in the limit \cite{Wu:TWC:2003}:
\begin{equation}
\label{eq:ECStandard}
\text{EC} = -\frac{\Lambda (-\theta)}{\theta}=-\lim_{t \to \infty} \frac{1}{\theta t} \log E (e^{-\theta S(t)}) \; [\text{bits/slot}]
\end{equation}
where $E(.)$ is the expectation operator and $S(t)=\sum_{k=1}^t s(k)$, with $s(k)$ as the channel service (i.e., number of bits delivered) during slot $k$. $\theta > 0$ is known as the QoS exponent; $\theta \to 0$ implies delay-tolerant communication, while $\theta \to \infty$ implies delay-limited communication.

\section{Mode Selection}

\subsection{Binary Hypothesis Testing (BHT)}
The mode selection problem for the D2D link for uplink transmission is defined as the following BHT:
\begin{equation}
	\label{eq:H0H1}
	 \begin{cases} H_0: & \text{direct mode ($D_T$ $\to$ $D_R$)} \\
                                 H_1: & \text{cellular mode ($D_T$ $\to$ $eNB$ $\to$ $D_R$).} \end{cases}
\end{equation}

The proposed mode selection method utilizes the {\it pathloss} as the sole feature to construct the BHT. Thus, the mode selection decides the direct mode when the pathloss of the D2D link is lesser than the pathloss of the $D_T\to$ eNB link, and vice versa. However, note that one could also utilize other physical-layer features, e.g., distance, received signal strength, signal-to-noise ratio (SNR), and instantaneous channel capacity \cite{mahmood2013mode}.

Let $\hat{L}$ denote the noisy measurement of $L$, the pathloss. Assume that $\hat{L} \sim \mathcal{N} (L,\sigma^2)$, where $\sigma^2$ is the variance of the estimation error. Also, let $T=\hat{L}_{d}-\hat{L}_{c,1}$ be the test statistic, where $\hat{L}_d$ ($\hat{L}_{c,1}$) represents the pathloss measurement on D2D ($D_T\to$ eNB) link. Note that $T|H_0\sim \mathcal{N}(-m_{T},\sigma_T^2)$, and $T|H_1\sim \mathcal{N}(m_{T},\sigma_T^2)$, where $m_{T}={L_d-{L}_{c,1}}$, and $\sigma_T^2=2\sigma^2$. Without loss of generality, let $m_{T}>0$.

Let $\pi(0)$ and $\pi(1)$ represent the prior probability of $H_0$ and $H_1$, respectively. Then, the mode selection problem\footnote{Since this work utilizes the pathloss as the sole feature, the mode selection needs to be carried out once every few seconds (when the pathloss changes).} is formulated as the following binary hypothesis testing problem (basically, a log-likelihood ratio test):
\begin{equation}
\label{eq:llrt}
LLR=\log_e (\frac{p(T|H_1)}{p(T|H_0)}) \underset{H_0}{\overset{H_1}{\gtrless}} \log_e \delta \implies T \underset{H_0}{\overset{H_1}{\gtrless}} \eta
\end{equation}
where $\delta=\frac{\pi(0)}{\pi(1)}$ and $\eta=\log_e (\delta). \frac{\sigma_T^2}{2m_T}$. Note that for the case of equal priors, $\log_e \delta=0$ (and thus $\eta=0$, which implies that the BHT simply checks the sign of test statistic $T$).\footnote{Inline with prior work \cite{mahmood2013mode}, this work assumes that the mode selection is carried out at the eNB, and the outcome is broadcasted to the D2D candidate pair ($D_T,D_R$) on the downlink control channel.}

\subsection{Performance of the BHT}
The performance of the BHT is quantified via two error probabilities. The probability of type-I error is given as:
\begin{equation}
P_{e,1} = P(H_1|H_0) = P(T>\eta|H_0) = Q(\frac{\eta+m_T}{\sigma_T})
\end{equation}
where $Q(x)=\frac{1}{\sqrt{2\pi}} \int_x^\infty  e^{-\frac{t^2}{2}} dt$ is the complementary cumulative distribution function (CDF) of $\mathcal{N}(0,1)$. Similarly: $P_{e,2} = P(H_0|H_1) = P(T<\eta|H_1) = 1 - Q(\frac{\eta-m_T}{\sigma_T})$.

Additionally, the Kullback-Leibler divergence (KLD) is a measure of how reliable the feature measurements are (and thus, the BHT). The KLD $D(p(T|H_1)||p(T|H_0))$ is given as: $D = \int_{-\infty}^{\infty} p(T|H_1) \log(\frac{p(T|H_1)}{p(T|H_0)}) dT = \frac{m_T^2}{\sigma_T^2}$.

\section{Effective Capacity Analysis}
\label{sec:EH}

We assume that channel state information at transmitter (CSIT) is not available at $D_T$. Thus, $D_T$ transmits data to $D_R$ at a fixed rate $r$ (bits/sec), and with a fixed average transmit power $\bar{P}$. In short, due to mode selection and no CSIT at $D_T$, the D2D link could be modelled as a Markov process.

\subsection{Markov Chain Modelling of Overlay D2D}
\label{subsec:MC}

Let $C_d(k), C_c(k)$ represent the instantaneous channel capacities (bits/sec) of the D2D link for the two hypotheses (direct mode, cellular mode), during slot $k$. When $r < C_d(k)$ ($r < C_c(k)$), then the direct link (cellular link) conveys $r$ bits/sec and is in ON condition. Otherwise, the D2D link conveys 0 bits/sec and is in OFF condition.\footnote{When the D2D link is in OFF condition, the bits need to be retransmitted, e.g., using the automatic repeat request mechanism.} This leads to the following four-state Markov process:

$s_1$: $H_0$ \& ON $\implies$ direct mode \& $r<C_d(k)$,

$s_2$: $H_0$ \& OFF $\implies$ direct mode \& $r>C_d(k)$,

$s_3$: $H_1$ \& ON $\implies$ cellular mode \& $r<C_c(k)$,

$s_4$: $H_1$ \& OFF $\implies$ cellular mode \& $r>C_c(k)$,

We now compute the two channel capacities as follows:
 \begin{equation}
 \label{eq:Cd}
{C_d}(k) = B \log_2 (1+\frac{\bar{P} |h_d(k)|^2}{L_{d} N_0}) = B \log_2 (1+\gamma_d(k))
\end{equation}
\begin{equation}
\label{eq:Cc}
\begin{split}
{C_c}(k) &= \frac{1}{2} \min \{C_{ul}(k),C_{dl}(k)\} \\
&= \frac{1}{2} B \log_2 (1+\min \{ \frac{\bar{P} |h_{c,1}(k)|^2}{L_{c,1} N_0}, \frac{\bar{P}_{eNB} |h_{c,2}(k)|^2}{L_{c,2} N_0} \}) \\
&= \frac{1}{2} B \log_2 (1+\min \{ \gamma_{ul}, \gamma_{dl} \}) = \frac{1}{2} B \log_2 (1+\gamma_c)
\end{split}
\end{equation}
where $C_{ul}$ and $\gamma_{ul}$ are the capacity and SNR of the uplink ($D_T\to eNB$) channel, respectively;  $C_{dl}$ and $\gamma_{dl}$ are the capacity and SNR of the downlink ($eNB\to D_R$) channel, respectively; $\gamma_{c}=\min \{ \gamma_{ul}, \gamma_{dl} \}$ is the net SNR of the cellular ($D_T\to eNB \to D_R$) link; $\gamma_{d}$, $h_d$, $L_d$ are the SNR, channel coefficient, and pathloss of the direct ($D_T\to D_R$) link, respectively; $h_{c,1}$ and $L_{c,1}$ are the channel coefficient and pathloss between $D_T$ and eNB, respectively; $h_{c,2}$ and $L_{c,2}$ are the channel coefficient and pathloss between eNB and $D_R$, respectively; $N_0$ is the noise variance at the eNB and $D_R$; $B$ is the bandwidth; and $\bar{P}_{eNB}$ is the average transmit power of the eNB. The pre-log factor of 1/2 in (\ref{eq:Cc}) is due to the fact that D2D communication in cellular mode consumes two slots.

For the states $s_1,s_2,s_3,s_4$ defined earlier, $p_{i,j}=[\mathbf{P}]_{i,j}$ is the transition probability from state $i$ to state $j$, with $\mathbf{P}$ as the transition probability matrix. The state of the D2D link changes after time $\tau$ (due to block fading). Next, we compute the state transition probabilities, starting with.
\begin{equation}
p_{1,1} = P\{  H_0(k) \; \& \; r<C_d(k)  |  H_0(k-1) \; \& \; r<C_d(k-1)  \}
\end{equation}
This can be equivalently expressed as:
\begin{equation}
p_{1,1} = P\{  H_0(k) \; \& \; \gamma_d(k)>\gamma_{req}  |  H_0(k-1) \; \& \; \gamma_d(k-1)>\gamma_{req}  \}
\end{equation}
where $\gamma_{req}=2^{r/B}-1$. Since the mode selection process $\{T\}_k$ is independent of the fading process $\{\gamma_d\}_k$, we can write:
\begin{equation}
p_{1,1} = P\{  H_0(k)|H_0(k-1) \} P\{ \gamma_d(k)>\gamma_{req}|\gamma_d(k-1)>\gamma_{req}  \}.
\end{equation}
Furthermore, we observe that both mode selection process $\{T\}_k$ and fading process $\{\gamma_d\}_k$ are memoryless, as they change independently from one slot to another. In other words, $P(H_0(k)|H_{\nu}(k-1))=P(H_0(k))$ for $\nu \in \{0,1\}$, and $P(\gamma_d(k)|\gamma_d(k-1))=P(\gamma_d(k))$. Therefore,
\begin{equation}
\label{eq:p11}
p_{1,1} = P(H_0(k)) P(\gamma_d(k)>\gamma_{req})
\end{equation}
where $P(H_0(k))=P(H_0|H_0)\pi(0)+P(H_0|H_1)\pi(1)$; $P(H_0|H_0) = 1 - Q(\frac{\eta+m_T}{\sigma_T})$. Since the SNR $\gamma_d(k)$ is exponentially distributed, then $P(\gamma_d(k)>\gamma_{req}) = 1 - P(\gamma_d(k)<\gamma_{req}) = e^{-\gamma_{req}/E(\gamma_d(k))}$, where $E(\gamma_d(k))=\frac{\bar{P}}{L_{d} N_0}$. Now, one can see that the transition probability $p_{1,1}$ does not depend on the original state. Therefore $p_{i,1} = p_1$.
Similarly,
\begin{equation}
\label{eq:p2p3p4}
\begin{split}
p_{i,2} = p_2 = P(H_0(k)) P(\gamma_d(k)<\gamma_{req}) \\
p_{i,3} = p_3 = P(H_1(k)) P(\gamma_c(k)>\gamma_{req}) \\
p_{i,4} = p_4 = P(H_1(k)) P(\gamma_c(k)<\gamma_{req}) \\
\end{split}
\end{equation}
where $P(\gamma_d(k)<\gamma_{req}) = 1-e^{-\gamma_{req}/E(\gamma_d(k))}$ and $P(H_1(k))=P(H_1|H_0)\pi(0)+P(H_1|H_1)\pi(1)$, with $P(H_1|H_1) = Q(\frac{\eta-m_T}{\sigma_T})$. Note that $\gamma_c(k)$ is an exponentially distributed random variable (R.V.) as well (because minimum of two exponentially distributed R.V.s is also an exponential R.V.). Thus, $P(\gamma_c(k)>\gamma_{req}) = e^{-\gamma_{req}/E(\gamma_c(k))}$, where $E(\gamma_c(k))=\frac{E[\gamma_{ul}]E[\gamma_{dl}]}{E[\gamma_{ul}]+E[\gamma_{dl}]}$, with $E[\gamma_{ul}]=\frac{\bar{P}}{L_{c,1} N_0}$ and $E[\gamma_{dl}]=\frac{\bar{P}_{eNB}}{L_{c,2} N_0}$. Finally, $P(\gamma_c(k)<\gamma_{req}) = 1-e^{-\gamma_{req}/E(\gamma_c(k))}$. With this, each row of $\mathbf{P}$ becomes: $\mathbf{p}_i=[p_1, p_2, p_3, p_4]$. Note that $\mathbf{P}$ has rank 1 due to identical rows.

\subsection{Effective Capacity of Overlay D2D}

We utilize the following result \cite{Chang:TNCS:2012}:
\begin{equation}
  \frac{\Lambda (\theta)}{\theta}=\frac{1}{\theta}\log_e sp(\mathbf{\Phi}(\theta)\mathbf{P})
\end{equation}
which states that for a Markovian service process $S(t)$, the log-MGF is given as $sp(\mathbf{\Phi}(\theta)\mathbf{P})$. Here, $sp(\mathbf{A})$ is the spectral radius of matrix $\mathbf{A}$ and $\mathbf{\Phi}(\theta)$ is a diagonal matrix containing the MGFs of the processes in the four states. Since $s(k)=r\tau$ bits for states $s_1,s_3$ and $s(k)=0$ bits for states $s_2,s_4$, the MGFs of the four states are $e^{\theta r \tau}$, 1, $e^{\theta r \tau}$, 1, respectively. Thus, $\mathbf{\Phi}(\theta) = \text{diag}(e^{\theta r\tau}, 1, e^{\theta r\tau}, 1)$. Since $\mathbf{\Phi}(\theta)\mathbf{P}$ is a matrix of unit-rank, finding its spectral radius is equivalent to finding its trace. Thus, $sp(\mathbf{\Phi}(\theta)\mathbf{P}) = (p_1+p_3)e^{\theta r\tau}+p_2+p_4$. Thus, the effective capacity (bits/sec) is:
\begin{equation}
\label{eq:EC}
\text{EC} = \frac{1}{\tau} \big[\frac{-1}{\theta} \log_e ((p_1+p_3)e^{-\theta r\tau}+p_2+p_4)\big].
\end{equation}
Note that the eNB could further compute the optimal rate $r^*$ as: $r^*=\arg \max_{r>0} EC$ and convey it to $D_T$. Moreover, $r^*$ is recomputed whenever the pathloss of the D2D link or ($D_T\to$ eNB) link changes (say, due to mobility of the D2D pair). Finally, when the D2D link and ($D_T\to$ eNB) link both experience more or less the same pathloss (i.e., $m_T\to 0$), mode selection collapses, and thus, Eq. (\ref{eq:EC}) does not hold.

{\bf Remark 1.} The cellular mode for D2D implies two-hop communication, and thus, two queues (at $D_T$ and eNB). To this end, we assume that the problem of overflow of (finite-sized) queue (due to backlog of bits due to transmission errors) arises at $D_T$ only. That is, we assume that the eNB knows the perfect CSI $h_{c,2}$, has infinite-sized queue (memory), and $\bar{P}_{eNB}>\bar{P}$; therefore, no backlog develops at the eNB. Under this setting, the definition of EC in (\ref{eq:ECStandard}) remains valid \cite{Lateef:TC:2009}.

\subsection{Effective Capacity of Underlay D2D}
In the underlay scenario, $D_R$ observes interference from $U_T$; therefore, we compute the signal-to-interference-plus-noise ratio (SINR) at $D_R$ in order to compute the two channel capacities $C_d(k), C_c(k)$ in (\ref{eq:Cd}), (\ref{eq:Cc}). When the D2D link operates in direct mode, the SINR is given as: $\Gamma_d(k)=\frac{\bar{P} |h_d(k)|^2/L_d}{I_d+N_0}$, where $I_d=\frac{\bar{P}_{U_T} |h_{U_T,D_R}(k)|^2}{L_{U_T,D_R}}$. Here, $h_{U_T,D_R}$ and $L_{U_T,D_R}$ represent the channel coefficient and pathloss between $U_T$ and $D_R$, and $\bar{P}_{U_T}$ is the average transmit power of $U_T$. When the D2D link operates in cellular mode, the SINRs on uplink and downlink are respectively given as: $\Gamma_{ul}(k)=\frac{\bar{P} |h_{c,1}(k)|^2/L_{c,1}}{I_{c,1}+N_0}$ and $\Gamma_{dl}(k)=\frac{\bar{P}_{eNB} |h_{c,2}(k)|^2/L_{c,2}}{I_{d}+N_0}$, where $I_{c,1}=\frac{\bar{P}_{U_T} |h_{U_T,eNB}(k)|^2}{L_{U_T,eNB}}$. Here, $h_{U_T,eNB}$ and $L_{U_T,eNB}$ represent the channel coefficient and pathloss between $U_T$ and eNB. Note that the computation of EC in (\ref{eq:EC}) requires re-computation of the four probabilities $p_1, p_2, p_3, p_4$. To this end, we consider an interference-limited scenario whereby we neglect noise to obtain the following signal-to-interference ratio (SIR) expressions: $\Upsilon_d=P_d^{(r)}/I_d$; $\Upsilon_{ul}=P_{c,1}^{(r)}/I_{c,1}$; $\Upsilon_{dl}=P_{c,2}^{(r)}/I_{d}$. Here, $P_d^{(r)}=\bar{P}|h_d|^2/L_d$; $P_{c,1}^{(r)}=\bar{P}|h_{c,1}|^2/L_{c,1}$; $P_{c,2}^{(r)}=\bar{P}_{eNB}|h_{c,2}|^2/L_{c,2}$. Observe that $P_d^{(r)}\sim \exp(\alpha)$ and $I_d\sim \exp(\beta)$, where $\alpha=L_d/\bar{P}$ and $\beta=L_{U_T,D_R}/\bar{P}_{U_T}$. Then, the CDF of $\Upsilon_d$ is: $P(\Upsilon_d<z)=\frac{\alpha}{\alpha+\beta/z}$. Then, $P(\Upsilon_d(k)>\gamma_{req})=1-\frac{\alpha}{\alpha+\beta/\gamma_{req}}$ and $P(\Upsilon_d(k)<\gamma_{req})=\frac{\alpha}{\alpha+\beta/\gamma_{req}}$, which allows us to compute $p_1$ and $p_2$ in (\ref{eq:p11}), (\ref{eq:p2p3p4}). As for $p_3$, $p_4$, let $\Upsilon_c=\min \{\Upsilon_{ul},\Upsilon_{dl}\}$. Also, observe that $P_{c,1}^{(r)}\sim \exp(\xi)$, $I_{c,1}\sim \exp(\zeta)$, and $P_{c,2}^{(r)}\sim \exp(\nu)$, where $\xi=L_{c,1}/\bar{P}$, $\zeta=L_{U_T,eNB}/\bar{P}_{U_T}$, and $\nu=L_{c,2}/\bar{P}_{eNB}$. Since $\Upsilon_{ul}$ and $\Upsilon_{dl}$ are independent R.V.s, the CDF of $\Upsilon_c$ is: $P(\Upsilon_c<z)=\frac{\xi}{\xi+\zeta/z}+\frac{\nu}{\nu+\beta/z}-\frac{\xi}{\xi+\zeta/z}\times \frac{\nu}{\nu+\beta/z}$. Thus, $P(\Upsilon_c(k)>\gamma_{req})=1-\big(\frac{\xi}{\xi+\zeta/\gamma_{req}}+\frac{\nu}{\nu+\beta/\gamma_{req}}-\frac{\xi}{\xi+\zeta/\gamma_{req}}\times \frac{\nu}{\nu+\beta/\gamma_{req}}\big)$ and $P(\Upsilon_c(k)<\gamma_{req})=\frac{\xi}{\xi+\zeta/\gamma_{req}}+\frac{\nu}{\nu+\beta/\gamma_{req}}-\frac{\xi}{\xi+\zeta/\gamma_{req}}\times \frac{\nu}{\nu+\beta/\gamma_{req}}$, which allows us to compute $p_3$ and $p_4$ in (\ref{eq:p11}), (\ref{eq:p2p3p4}).

\section{Numerical Results}
\label{sec:results}

A cell with radius of 700 m is considered, and the two pairs of UEs are placed inside the cell according to the uniform distribution. To obtain pathloss measurements for mode selection, the pathloss model in \cite{Mahboob:Globecom:2015} was used: $L(d) = 36.3 + 37.6 \log_{10}(d)$. We set $B=10$ kHz, and $\tau=0.1$ sec. Finally, we set $\pi(0)=\pi(1)=0.5$ which implies $P_{e,1}=P_{e,2}$.

\begin{figure}[ht]
\begin{center}
	\includegraphics{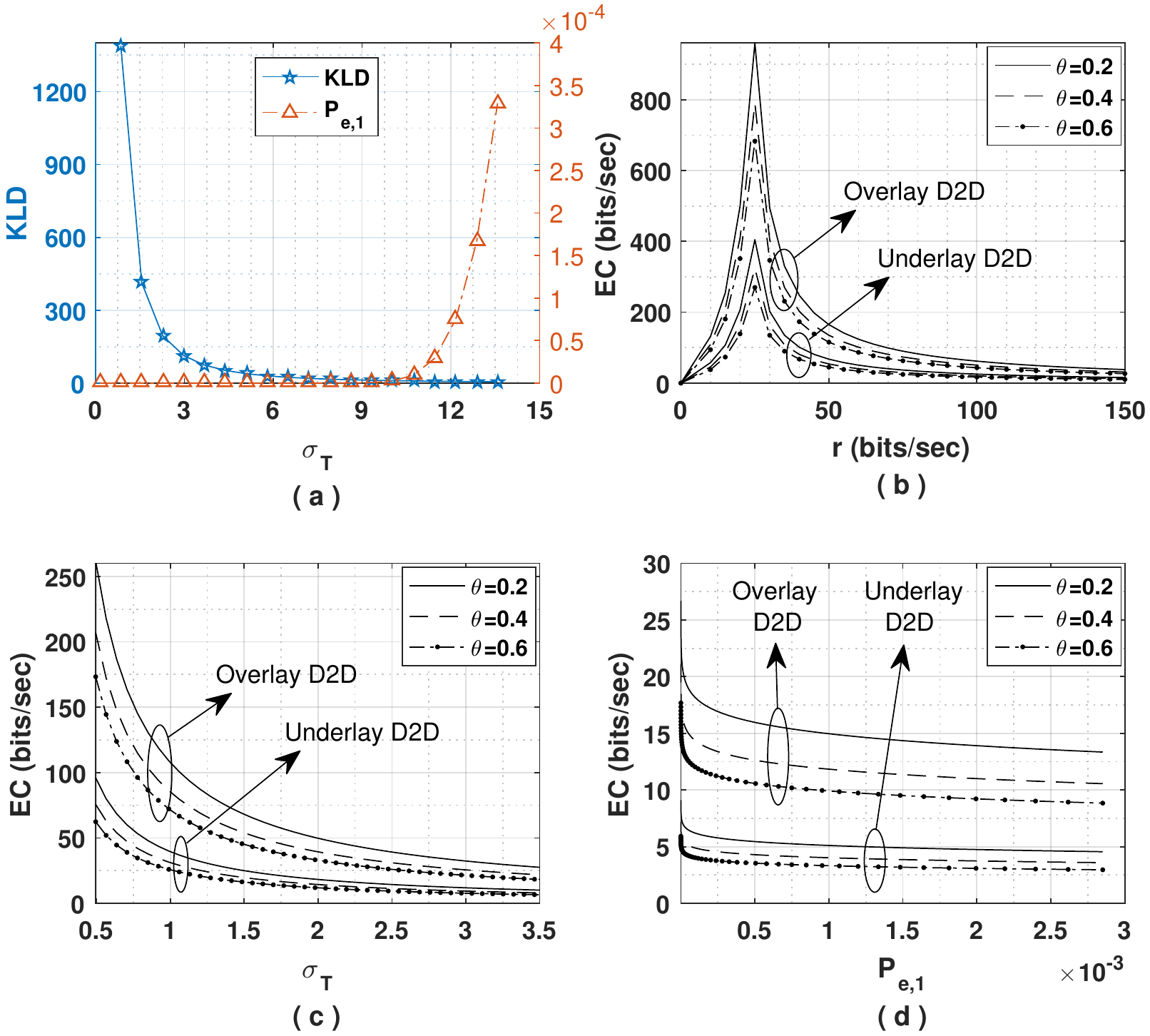}
\caption{(a) Impact of $\sigma_T$ on mode selection, (b) Exhaustive search to compute the optimal rate $r^*$ maximizing the EC, (c) Impact of $\sigma_T$ on the EC, (d) Impact of $P_{e,1}$ on the EC.}
\label{fig:ecsigma}
\end{center}
\end{figure}

Fig. \ref{fig:ecsigma} (a) demonstrates that the mode selection could sustain pretty large estimation errors, i.e., $P_{e,1}$ does not grow until $\sigma_T>10.5$. Next, Fig. \ref{fig:ecsigma} (b) shows results of the exhaustive search for the optimal rate maximizing the EC, which is $r^*=25$ bits/sec (regardless of the QoS constraint level $\theta$). Fig. \ref{fig:ecsigma} (c) shows that the EC (as well as the gain of overlay D2D over underlay D2D) decreases exponentially fast; in fact, EC becomes nearly zero for $\sigma_T>3.5$. This suggests the need to design efficient estimators for the pathloss (the feature utilized by the mode selection) meeting the Cramer-Rao bound (CRB). The findings of Fig. \ref{fig:ecsigma} (d) are the same as Fig. \ref{fig:ecsigma} (c). Finally, Figs. \ref{fig:ecsigma} (b)-(d) all show that more strict QoS requirements lead to lesser EC and vice versa, as expected.


\section{Conclusion}
\label{sec:conclusion}

This work investigated the impact of mode selection on the EC of a D2D link. The derived analytical expression of the EC, as well as simulation results, reveal that the EC decreases exponentially fast as the pathloss measurements become more noisy. Thus, designing efficient estimators (of the features considered by the mode selection) meeting the CRB is of paramount importance. Additionally, for fixed $\sigma_T$ and $\theta$, the expression in Eq. (\ref{eq:EC}) further allows us to compute the optimal (fixed) rate $r^*$ which maximizes the EC.

\appendices
\footnotesize{
\bibliographystyle{IEEEtran}

}

\vfill\break

\end{document}